# Interplay between Kondo effect and Ruderman-Kittel-Kasuya-Yosida interaction

Henning Prüser[1], Piet E. Dargel[2], Mohammed Bouhassoune[3], Rainer G. Ulbrich[1], Thomas Pruschke[2], Samir Lounis[3] and Martin Wenderoth[1]

[1]4. Physikalisches Institut, Georg-August-Universität Göttingen, Friedrich-Hund-Platz 1, 37077 Göttingen, Germany

[2]Institut für Theoretische Physik, Georg-August-Universität Göttingen, Friedrich-Hund-Platz 1, 37077 Göttingen, Germany

[3]Peter Grünberg Institut and Institute for Advanced Simulation, Forschungszentrum Jülich & JARA, 52425 Jülich, Germany

**The interplay between the Ruderman-Kittel-Kasuya-Yosida (RKKY) interaction and the Kondo effect is expected to provide the driving force for the emergence of many phenomena in strongly correlated electron materials. Two magnetic impurities in a metal are the smallest possible system containing all these ingredients and define a bottom up approach towards a long term understanding of concentrated / dense systems. Here we report on the experimental and theoretical investigation of iron dimers buried below a Cu(100) surface by means of low temperature scanning tunnelling spectroscopy (STS) combined with density functional theory (DFT) and numerical renormalization group (NRG) calculations. The Kondo effect, in particular the width of the Abrikosov-Suhl resonance, is strongly altered or even suppressed due to magnetic coupling between the impurities. It oscillates as function of dimer separation revealing that it is related to the RKKY interaction mediated by the conduction electrons. Simulations based on density functional theory support this concept showing the same oscillation period and trends in the coupling strength as found in the experiment.**

Magnetic atoms embedded in a non-magnetic host metal have been intensively studied in solid state physics during the last decades. At low temperature these systems can exhibit strong electronic correlations that give rise to various fascinating phenomena beyond the single particle picture [1]. The magnetic impurities, elements with partially filled d-, or f-orbitals, have localized spins that interact with the surrounding conduction electrons. Below a characteristic temperature $T_k$, the coupling between conduction electrons and the localized spins leads to the Kondo effect, a collective screening of the impurity spins. Since the impurities are coupled to the same conduction electrons a magnetic interaction between impurities leading to magnetic ordering may arise. This indirect exchange interaction is well-known as Rudermann-Kittel-Kasuya-Yosida (RKKY) interaction [2-4]. Many phenomena in strongly correlated electron physics, especially for heavy-fermion metals like unconventional superconductivity or quantum critical behaviour are attributed to the competition of RKKY interaction with Kondo physics [5]. While single-impurity Kondo physics, where the impurities can be treated individually, has been studied for a long time in macroscopic measurements [6] and more recently in nanostructures [7-18] a detailed knowledge of the transition towards a systems of interacting impurities is still lacking.

Although, the concept of spatially extended indirect exchange in the two-impurity Kondo problem is well established in theoretical research, experimental studies report a rather short ranged interaction between Kondo impurities. Scanning tunnelling spectroscopy (STS) of two magnetic atoms adsorbed on a noble metal surface found a significant change of Kondo features only for small dimer distances, e.g. a few angstroms [19-22]. Attaching two magnetic atoms at the ends of a non-magnetic chain may enhance the range of the magnetic exchange interaction [23], but structural relaxation in these nanostructures make it difficult to compare different chain lengths with each other. Furthermore, this approach is limited towards certain directions and hence gives no access to all possible discrete separations on a surface. Two interacting Kondo impurities can also be realized by two coupled quantum dots [24, 25]. However, due to geometrical constraints, the spatial dependence of the interaction strength cannot be studied in these experiments. Recent studies showed, that the short distance regime can be experimentally explored by atomic point contacts [26].

Here we address the transition from a single Kondo impurity towards a coupled system and investigate experimentally as well as theoretically the interaction between two magnetic impurities. In particular, the study of spatial extension and strength of the RKKY interaction in a bulk system, as reported here, may help to develop a broader understanding of strongly correlated electron systems.

**Results**

**Fe dimers below a copper surface.** In our experiment we investigate single isolated magnetic Fe dimers buried below the Cu(100) surface with a low-temperature STM (see Methods). We have chosen Fe in copper because of its low Kondo temperature [16]. Furthermore, in this work we focus on dimers located in the fifth copper monolayer below the surface [17] due to two reasons. Firstly, we are interested in bulk impurities and for dimers located within the first monolayers the presence of the surface may strongly influence the magnetic interaction. Secondly, for much deeper impurity positions the energy dependence of the band structure itself may cause strong spectroscopic signatures around the Fermi energy which make it hard to extract microscopic information of the Kondo system [27]. In our experiment dimers can be identified by their topographic signature. When two Fe atoms are very close to each other their interference pattern starts to overlap. The dimer configuration within the atomic lattice can be extracted by analysing the topography (see Supplementary Note 1 and Supplementary Fig. S1).

As a first striking example Fig. 1 depicts that two adjacent Fe impurities which are separated by R = 0.57 nm show a considerably different spectral signature compared to a single impurity. For a tip position direct at the centre of the interference pattern of the dimer a narrow peak is found in the background subtracted differential conductance $\Delta dI/dV$ (see Methods). For equivalent spatial positions on the interference pattern of the dimer and of the monomer (Fig. 1a and Fig. 1b respectively) nearly the same line shape is observed (Fig. 1c). A closer look at the spectroscopic data, however reveals two main differences. At first, the signature found in the centre of the standing-wave pattern of the dimer (purple dots) is twice as high as the signature with a tip position at the border (blue dots). This feature can be explained by the vicinity of the second Fe impurity which induces the same signature at the centre as the first one. More significantly, the resonance width for the dimer is broadened compared to a monomer (blue and green dots), while the line shape as well as the amplitude is nearly the same for both spectra. Up to now there exists no fit formula or model for a magnetic dimer in an STM experiment taking into account the electron propagation of the material and all different scenarios which may occur in the system. Our many-body calculations (see below), however suggest that the same fit formula as for the single impurity case also describes the spectral signature of two interacting impurities (see Supplementary Fig. S2 and Supplementary Note 2). We use a fit formula based on the model of Újsághy and colleagues [28], but replace the Lorentzian form of the Kondo resonance by a phenomenological form of the Kondo resonance proposed by Frota and co-workers [29, 30]. The final expression correctly describes Kondo features such as line-shape [31] and more dramatically the phase shift caused by the resonance fairly well in both experiment and many-body calculations [16, 17, 32, 33]. By fitting the spectroscopic signal of the dimer the half width at half maximum (HWHM) $\Delta$ of the Kondo resonance of 16.5 ± 1.5 mV and 15.0 ± 3.6 mV is extracted with a tip position at the centre and at the border, respectively. Both values are nearly the same and slightly but significantly higher than the resonance width $\Delta$ = 12.2 ± 2.3 mV observed for the single Fe impurity. Please note, that the data for the monomer and dimer was measured with the same tip on the same sample, so that the observed differences in the spectroscopic signature are not related to tip properties.

**Different dimer configurations.** In order to obtain deeper insights into the interaction of two Kondo atoms with each other, an overview of different iron dimer configurations with their corresponding spectroscopic signature is presented in Fig. 2. All dimers (Fig. 2a-e) are located in the fifth monolayer and the interatomic distance R between adjacent iron atoms ranges from R = 0.51 nm to R = 1.02 nm (Fig. 2f-j). For the most compact dimer configuration no signature of the Kondo effect is observed (Fig. 2k, Supplementary Fig. S3 and Supplementary Fig. 4). Further details for this configuration are discussed in Supplementary Note 3. With increasing interatomic distance features at the Fermi energy due to the Kondo effect are found (Fig. 2l-o). All signatures can be fairly well described by the phenomenological fit formula (parameters given in Supplementary Table S1). The different configurations show a variation of the Kondo resonance width $\Delta$, depicted in Fig. 4a. All values are increased compared to a single iron impurity which exhibits a resonance width of $\Delta_0$ = 11.7 ± 3.0 mV. This value is obtained by averaging over different measurements with different tips and lateral tip-positions. Remarkably, the resonance width found for the different dimers is not a monotonic function of their interatomic distance. While for R = 0.57 nm and R = 0.72 nm (purple and blue configuration) the width of the Kondo resonance is slightly increased, a separation of R = 0.77 nm (green configuration) shows a value which is nearly two times larger compared to the single impurity value. For larger separation R = 1.02 nm (orange configuration) the resonance width decreases recovering nearly the single impurity reference value.

The experimental findings can be understood within the two-impurity Kondo model (TIKM), the simplest theoretical model to describe magnetic interactions between Kondo impurities [34-36]. The qualitative ground state properties of the TIKM can be derived by considering the two competing effects (see Figure 3a). On the one hand, the single impurity Kondo effect, which tries to screen each impurity spin independently, and on the other hand the magnetic interaction between the impurities, promoting a non-local correlation between the localized spins.

Spectral properties of the TIKM can be obtained for T = 0 K by using numerical renormalization group calculations [37]. Here, we use as model for the Cu host a free electron gas with Fermi wave vector $k_F$ and other model parameters such that the individual impurities are well in the Kondo regime (see Methods). No direct exchange is included, i.e. only RKKY generated spin-correlations are considered. Varying the inter-impurity distance R, we first observe in Fig. 3b the expected oscillation of the spin-spin correlation function, reflecting the corresponding behaviour of the RKKY interaction [36]. Note that the correlations decay rapidly on a scale of the Fermi wave length $\lambda_F$, and start out antiferromagnetic (AFM) as the impurities approach each other, becoming ferromagnetic (FM) as the separation drops below one quarter of the Fermi wave length. The corresponding spectral functions are collected in Fig. 3c. As expected, they resemble the spectra of independent impurities for large R, showing the particular hallmark of the many-body singlet state, a narrow resonance at the Fermi energy. With decreasing R, in the regime of antiferromagnetic correlations, we observe an enhanced width, which is most apparent from the inset to Fig. 3c, where the spectra are normalized to their value at the Fermi energy. Similar results were obtained recently by applying perturbation theory [38].

In general, starting from well separated impurities, one expect first an increase of the HWHM $\Delta$ of the Kondo resonance for antiferromagnetic correlations, showing a maximum for the configuration with the strongest antiferromagnetic coupling. For small distances ferromagnetic coupling takes over leading to a reduction of $\Delta$. This situation is similar to multi-orbital Kondo physics increasing the total spin of the system, which results in a strongly reduced Kondo scale [39, 40].

From the distances of the iron dimer configurations investigated in our experiment and the oscillatory behaviour of Kondo resonance width we can exclude the possibility of a direct exchange mechanism. More likely, the coupling results from the indirect RKKY interaction. First-principles calculations (see Methods) support this idea: the magnetic exchange interactions $J_{ex}$ estimated for the different dimer configurations are depicted in Fig. 4b. The observed width of the Kondo resonance and the calculated magnetic interaction strength have the same oscillation period. For the smallest separation with R = 0.51 nm (red configuration) an AFM exchange coupling with $J_{ex}$ = -21.8 meV is predicted. Here no Kondo signature around zero bias is found in the experiment. The AFM coupling between the localized impurity spins seems to be stronger than the coupling to the surrounding conduction electrons and hence the Kondo effect is suppressed.

A slightly weaker AFM exchange interaction of $J_{ex}$ = -15.2 meV is calculated for a distance R = 0.77 nm (green configuration), experimentally showing the strongest broadening. Configurations

R = 0.57 nm and R = 1.02 nm exhibiting smaller magnetic exchange interaction $J_{ex}$ and hence cause a smaller broadening. Following the simulation a FM coupling is expected for a separation of R = 0.72 nm. According to the TIKM this may lead to a reduction of the Kondo resonance width. Experimentally, an increased value is found. One possible explanation is that the FM exchange coupling is too weak to produce a triplet state at the experimental temperature of 6 K. As a consequence, the impurity spins are not fully screened which lead to an increased resonance width [41].

**Spatial dependence of the RKKY interaction.** A detailed analysis of our results reveals a strong directional dependence of the exchange interaction (see Supplementary Fig. S5). Both theory and experiment highlight that the [01$\bar{1}$] direction promotes the strongest magnetic coupling between impurities. Compare for example the red with the purple dimer configuration (or the green with the blue one). Here the distance between the iron atoms is nearly the same, but $J_{ex}$ is stronger for the dimer oriented along the [01$\bar{1}$] direction. This feature was also observed for magnetic dimers on a metal surface [42, 43] and may be explained by an aliasing between the periodicity of the lattice and the wave length of the RKKY oscillation. As the RKKY interaction is mediated by the conduction electrons also anisotropies in the band structure can cause a directional dependence. Copper is well known to show strong directional Friedel oscillations with wave vectors and amplitudes depending on the crystal direction [17, 44-47]. Indeed, the two-dimensional spectral function projected on the fifth layer below the surface (Fig. 4c) highlights the Fermi energy contours weighted with the electronic occupation. Interestingly a squared-like shape is obtained with flat regions perpendicular to the diagonals. This is the required ingredient to obtain a strong focusing effect of the electronic beams mediating the magnetic exchange interactions among the impurities. The shape obtained is reminiscent of the energy contour obtained in the bulk of copper. Interestingly, our simulations demonstrate that once the impurities are put on top of the surface or embedded in the surface, the magnetic exchange interactions decrease strongly since the corresponding bulk energy contours lose their intensity (see Supplementary Note 4 and Supplementary Fig. S7).

**Discussions**

Many of the experimental and theoretical results presented here for dimers located in the fifth monolayer are also obtained for dimers located in other monolayers, indicating that the RKKY interaction is only slightly affected by the presence of the surface (Fig. S.3). Our experimental findings are therefore not only relevant for artificially nanostructures but also may be relevant for the investigation of strongly correlated bulk systems including a crossover from local Kondo screening of the constituents to the formation of magnetic ordering. Being able to measure the properties of two magnetic bulk impurity is a prerequisite for the understanding and design of more complex structures. For example trimers or linear chains are promising model systems for a bottom up approach towards materials where the local moments are arranged in a periodic array, so-called Kondo lattices. Up to now, theoretical models describing these systems are mainly isotropic and mostly consider only nearest or next nearest neighbour magnetic interactions. This work illustrates that the RKKY interaction between two bulk Kondo impurities can be significantly large also for greater distances and furthermore shows strongly directional behaviour. For iron dimers in copper the RKKY interaction can increase the Kondo resonance width by a factor of two for a separation which is more than twice the length of the unit cell. As bulk impurities are well-known to show universal behaviour to a certain extent we expect that magnetic interactions at larger distances may be significant also in other systems.

Future experimental and theoretical work will focus on how multi-orbital Kondo physics and magnetic anisotropy affect the RKKY interaction. While for Fe in copper we do not see any significant deviations from a coupled spin 1/2 model, systems with a lower Kondo temperature may be strongly affected by spin-orbit coupling and crystal field splitting [13, 18]. Manganese or rare earth impurities are interesting candidates for such an experiment.

**Methods**

**Scanning tunnelling microscopy and spectroscopy.** The STM experiments presented here were performed using a home built low-temperature scanning tunnelling microscope operating at 6K at pressures below $5 \cdot 10^{-11}$ mbar. The Cu(100) single crystal substrate is cleaned by several cycles of argon bombardment and electron beam heating. A copper alloy with a small amount (~0.04%) of iron impurities is prepared through simultaneous deposition of copper and iron from two electron beam evaporators. We use electrochemically etched tungsten tips, prepared by annealing and argon ion bombardment. The performance of the tip is tuned by controlled voltage pulses and smooth tip-sample contacts. The x- and y-axis in the constant current topographies are aligned to the [010] and [001] crystallographic direction, respectively. Differential conductance measurements are obtained using a lock-in technique adding a modulation of 1.4 mV – 2.0 mV to the sample voltage at a frequency of 2190 Hz. The spectra are acquired by setting the tip height to give a set point current $I_{SP}$ at a set point voltage $V_{SP}$, holding the tip at this fixed position above the surface, and then sweeping the voltage while recording dI/dV. The spectra are not sensitive on the impedance of the tunnelling junction. Typical differential conductance dI/dV spectra measured with a tip position above the interference pattern of a Fe impurity reveal a strong signature around zero bias voltage which is attributed to the Kondo effect and cannot be seen in spectra of the bare Cu surface. In order to remove artefacts which originate from the tip, we normalize the dI/dV data around an impurity by subtracting from the measured signal the spectra of the bare surface $dI_0/dV$ far away from the impurity [48].

**Numerical renormalization group calculations.** Magnetic impurities in metals are conventionally described using a variant of the Anderson impurity model [1]. Here, we study the concrete model

$$\hat{H} = \sum_{\vec{k}\sigma} \epsilon_{\vec{k}} \hat{n}_{\vec{k}\sigma} + \sum_{i\sigma} \left( \epsilon_d + \frac{U}{2} \hat{n}_{i,-\sigma} \right) \hat{n}_{i\sigma} + \frac{V}{\sqrt{N}} \sum_{i\vec{k}\sigma} \left( e^{i\vec{k}\cdot\vec{R}_i} \hat{c}^\dagger_{\vec{k}\sigma} \hat{d}_{i\sigma} + h.c. \right)$$

where we used standard notation for the different terms. The two magnetic impurities are located at sites $R_i$, i = 1,2. Note that no explicit exchange between the impurities is included. To proceed, one introduces linear combinations, $\hat{d}_{\pm,\sigma} := \frac{1}{\sqrt{2}}(\hat{d}_{1\sigma} \pm \hat{d}_{2\sigma})$, for the impurity degrees of freedom. By this transformation one obtains a model where the states with even respectively odd parity couple via a hybridization function, $\Gamma_\pm(\omega) := \frac{V^2}{N}\sum_{\vec{k}}[1 \pm \cos(\vec{k}\cdot\vec{R}_{12})]\delta(\omega - \epsilon_{\vec{k}})$ , to the corresponding band-electron degrees of freedom [49]. Assuming a free electron gas in D = 3 for the band electrons, one can approximate this by $\Gamma_\pm(\omega) := \Gamma_0\left[1 \pm \frac{\sin(kR_{12})}{kR_{12}}\right]$ and use a linearized dispersion $\epsilon_k \approx D\left(\frac{k}{k_F} - 1\right)$ for the explicit evaluation. Note that the relevant parameter to control the impurity distance in the model then becomes $k_F R_{12}$.

Still, the model is a quantum impurity model which needs to be solved with a suitable numerical technique. We here employ the numerical renormalization group (NRG) [37], which allows us to solve the two-impurity model at T = 0 in the Kondo regime. To that extent we choose a value $U/(\pi\Gamma_0) = 3$ and $\epsilon_d = -U/2$. The NRG parameters were: Discretization parameter Λ=2.5, 3000 states per iteration kept with 100 NRG iterations in total. We calculated the spectral functions directly without employing the self-energy. Please note that the many-body calculations can be described by a phenomenological form of the Kondo resonance proposed by Frota and co-workers [29, 30], see Supplementary Fig. S2.

**Ab initio calculations.** The Korringa-Kohn-Rostoker Green function method (KKR) [50, 51] within the local-density approximation (LDA) of the density functional theory (DFT) is used to simulate and investigate the different Fe-dimers embedded underneath the surface of copper. The magnetic exchange interactions among Fe impurities were extracted using the frozen-potential approximation considering infinitesimal rotations of the magnetic moments [52]. As in the experiment, the iron impurities are located in the fifth monolayer below the Cu(100) surface. The magnetic exchange interaction $J_{ex}$ between two magnetic moments, has been extracted by mapping the *ab initio* calculations to the Heisenberg model

$$H = -J_{ex}\vec{e}_1 \cdot \vec{e}_2$$

The magnetic exchange interaction $J_{ex}$ describes the coupling between the localized impurity spins $\vec{S}_1$ and $\vec{S}_2$ with $\vec{e}_1$ and $\vec{e}_2$ being their unit vectors.

**Acknowledgement**

We acknowledge Kurt Schönhammer, Philipp Kloth, Ben Warner and Cyrus Hirjibehedin for stimulating discussions. This work was supported by the DFG through SFB 602 Project A3 and by the HGF-YIG Programme VH-NG-717 (Functional Nanoscale Structure Probe and Simulation Laboratory-Funsilab).

**Author contributions**

H.P, M.W. and R.G.U. planned the experiments. H.P. carried out the experiments and the data analysis. S.L. and M.B. did and analysed the first principles calculations. T.P. and P.E.D. did the NRG calculations. H.P. wrote the manuscript. All authors discussed the results and commented on the manuscript.

Corresponding author M.W. (email: mwender@gwdg.de)

**Figures**

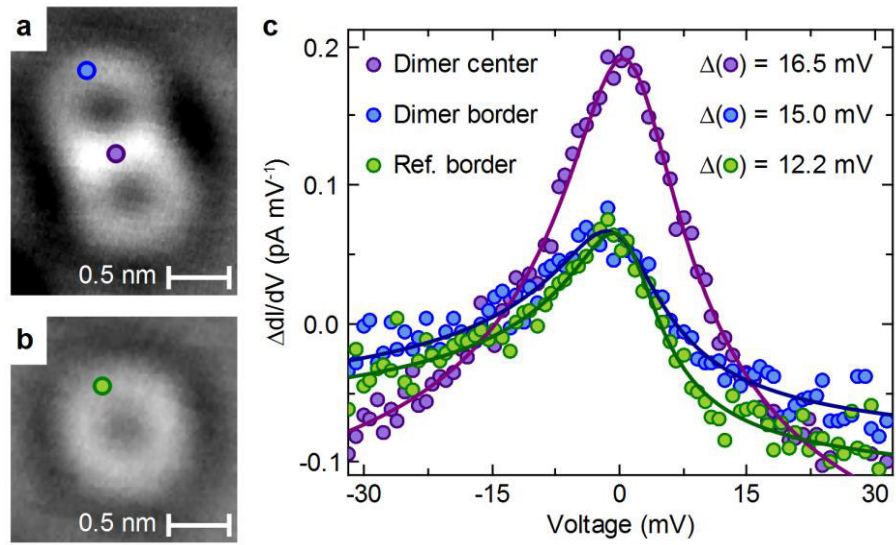

**Figure 1. Comparison of an iron dimer with an isolated single impurity located in the same monolayer. a**, STM constant current topographies of a dimer with a separation of R = 0.57 nm and **b**, a monomer. **c**, Single ΔdI/dV spectra for different lateral tip positions, marked in the topographies by circles. The solid curves show fits to the spectra. At the centre of the interference pattern of the dimer (purple dots) a peak like feature is observed. Compared to the monomer the half width at half maximum (HWHM) of the Kondo resonance Δ is increased. More clearly this effect can be seen in spectra with the same lateral tip positions (green and blue dots). The resonance width of the dimer is broadened, while the line shape as well as the amplitude is nearly the same for both spectra. Image parameters are (**a, b**) V = -31 mV, I = 0.2 nA, colour scale from -4.0 pm black to 5.5 pm white. Spectroscopy parameters are (**c**) $V_{SP}$ = -40 mV, $I_{SP}$ = 0.3 nA.

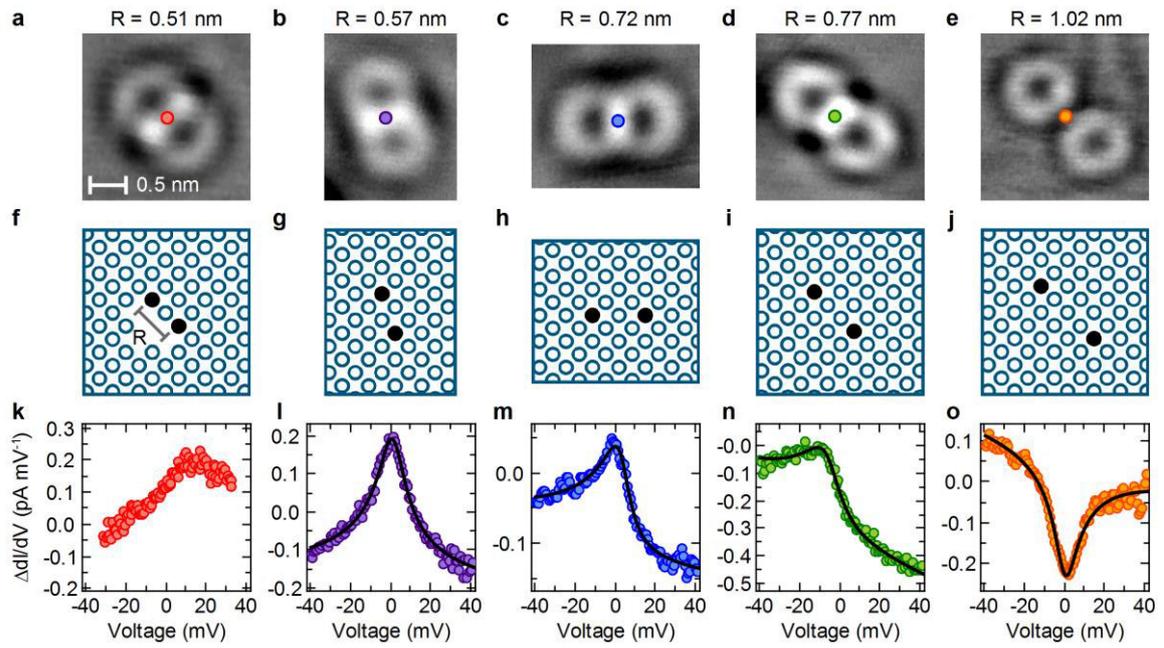

**Figure 2. Overview of different dimer configurations.** All dimers are situated in the fifth monolayer below the Cu(100) surface. **a-e**, Constant current topographies of different dimer configurations with increasing distance. **f-j**, Corresponding atomic configurations. **k-o**, Single ΔdI/dV spectra (dots) with a tip position at the centre of the standing-wave pattern. While for the smallest dimer configuration (**f**) no feature around zero bias is observed, all other configurations (**g-j**) show Kondo fingerprints. The solid black curves show fits to the spectra. Image parameters are (**a-e**) V = -31 mV, I = 0.2 nA, colour scale from -4.0 pm black to 5.5 pm white. Spectroscopy parameters are (**k**) $V_{SP}$ = -31 mV, $I_{SP}$ = 0.25 nA; (**l, n**) $V_{SP}$ = -40 mV, $I_{SP}$ = 0.3 nA; (**m, o**) $V_{SP}$ = -40 mV, $I_{SP}$ = 0.25 nA.

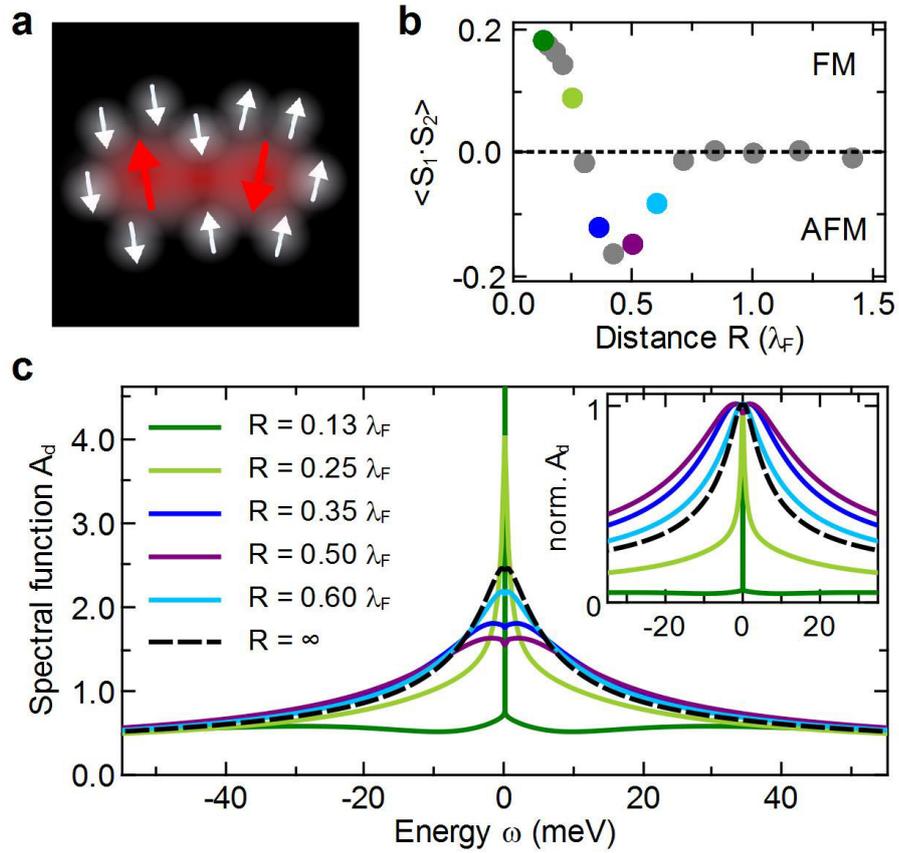

**Figure 3. Competition of Kondo effect and magnetic exchange interaction between two impurities. a**, Schematic view of the two-impurity Kondo model, illustrating the screening of the local moments (red arrows) by the itinerant conduction electrons (white arrows). Since both impurities are coupled to the same conduction electrons a magnetic exchange interaction $J_{ex}$ arises intrinsically between the two impurity spins ($S_1$ and $S_2$). **b**, The spin-spin correlation function of the localized moments shows an oscillation between ferromagnetic and antiferromagnetic coupling as function of their distance R (given in multiples of the Fermi wavelength $\lambda_F$). **c**, Evolution of the spectral function $A_d$ of the two-impurity Kondo model for different distances. The inset shows the normalized spectral function with respect to their value at the Fermi energy.

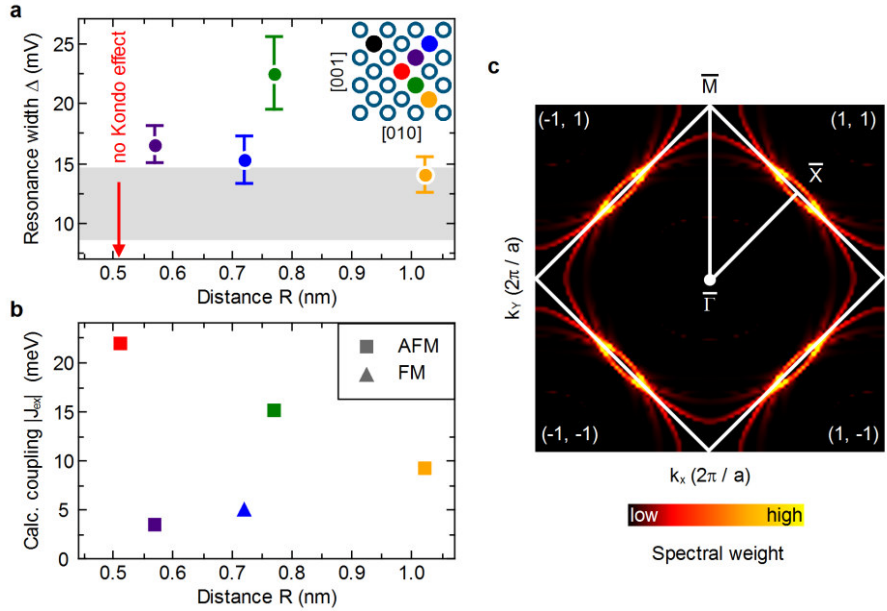

**Figure 4. Distance dependence and magnetic exchange coupling. a**, Observed Kondo resonance width $\Delta$ as function of the dimer separation R. The corresponding atomic configurations within the fifth monolayer are shown in the inset (black dot: position of the first Fe atom, colour dot: position of the second Fe atom). As reference the averaged value for a single isolated fifth monolayer Fe impurity $\Delta_0 = 11.7 \pm 3.0$ mV is highlighted in light grey. **b**, Absolute calculated values for the exchange interaction $|J_{ex}|$ between two iron atoms below the Cu(100) surface. Ferromagnetic configurations are marked as triangles while antiferromagnetic configurations as squares. **c**, Bloch spectral function at the Fermi energy projected on the layer in which the impurities are embedded. The two-dimensional Fermi energy contours are of almost squared like-shape with flat regions along the directions which is reminiscent of the bulk Fermi surface of copper projected on the plane perpendicular to the (100) direction.

# Supplementary Information

# Interplay between Kondo effect and Ruderman-Kittel-Kasuya-Yosida interaction

Henning Prüser[1], Piet E. Dargel[2], Mohammed Bouhassoune[3], Rainer G. Ulbrich[1],
Thomas Pruschke[2], Samir Lounis[3] and Martin Wenderoth[1]

[1]4. Physikalisches Institut, Georg-August-Universität Göttingen, Friedrich-Hund-Platz 1, 37077 Göttingen, Germany

[2]Institut für Theoretische Physik, Georg-August-Universität Göttingen, Friedrich-Hund-Platz 1, 37077 Göttingen, Germany

[3]Peter Grünberg Institut and Institute for Advanced Simulation, Forschungszentrum Jülich & JARA, 52425 Jülich, Germany

**Supplementary Figures**

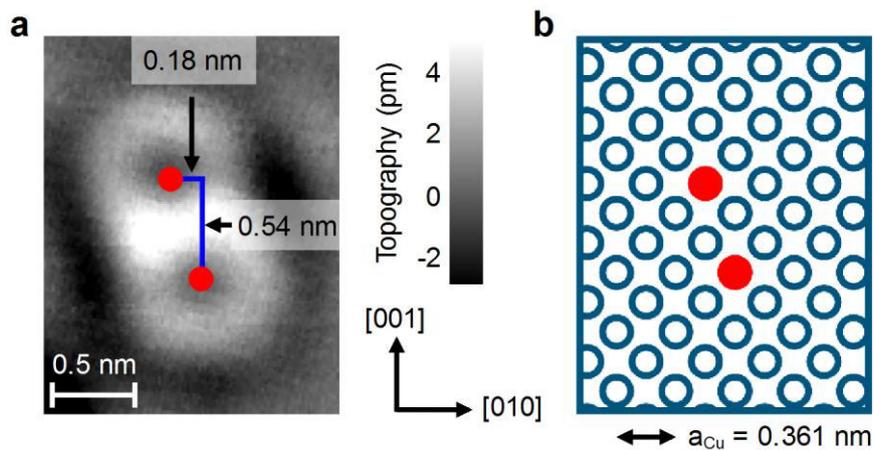

**Supplementary Figure S1. Extracting the atomic configuration of two iron atoms close to each other. a**, STM constant current topography (1.8 nm × 2.3 nm) showing two neighbouring fifth monolayer Fe impurities. The centre of each standing wave pattern, marked by the red dots, is displaced 0.18 nm and 0.54 nm, respectively, along the [010] and [001] direction. **b**, Corresponding atomic lattice, the position of the Fe impurities is marked in red. Image parameters are (**a**) V = -31 mV, I = 0.2 nA.

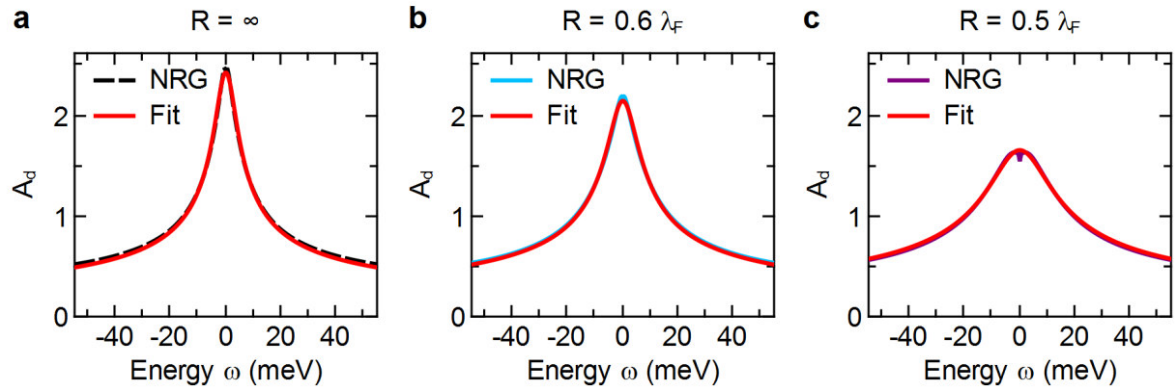

**Supplementary Figure S2. Calculated spectral function $A_d$ of the two-impurity Kondo model.** For a dimer separation of $R = \infty$ (a), $R = 0.6\,\lambda_F$ (b) and $R = 0.5\,\lambda_F$ (c). The calculated spectra can be fairly good described by the phenomenological fit formula obtained by Frota (red curves).

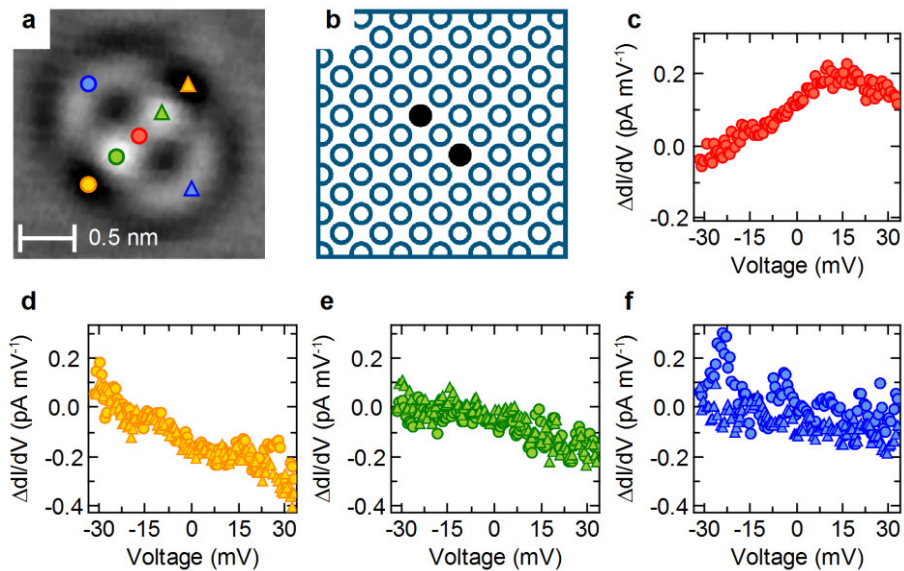

**Supplementary Figure S3. Disappearance of the Kondo effect for a compact dimer configuration. a**, STM constant current topography of the dimer with R = 0.51 nm and **b**, corresponding atomic configuration. **c-f**, Single $\Delta dI/dV$ spectra for different lateral tip positions, marked in the topography. All spectra are featureless around zero bias. The Kondo resonance observed for a single impurity is no longer present for this dimer configuration. Image parameters are (**a**) V = -31 mV, I = 0.2 nA. Spectroscopy parameters are (**c-f**) $V_{SP}$ = -31 mV, $I_{SP}$ = 0.25 nA.

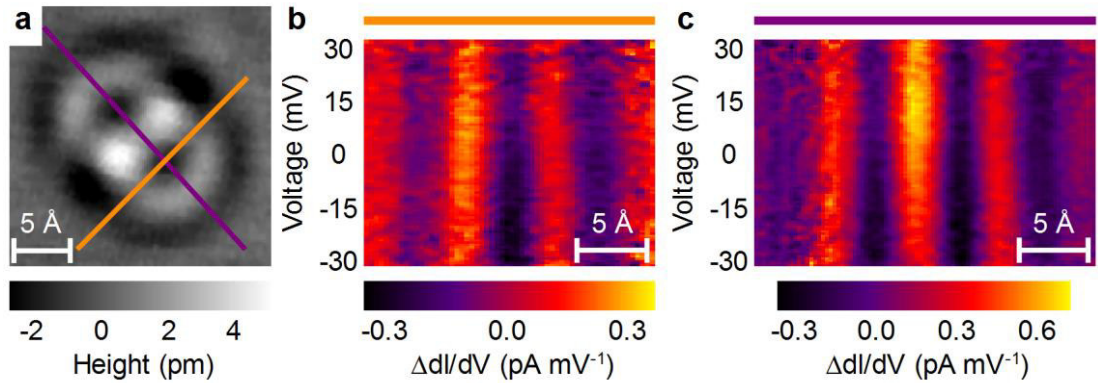

**Supplementary Figure S4. Scattering behaviour of a compact dimer configuration. a**, STM constant current topography of the dimer with R = 0.51 nm. **b**, ΔdI/dV signal as function of applied sample bias voltage and one spatial coordinate across the standing-wave pattern of the dimer (indicated by the orange line in **a**). **c**, Section of the differential conductance along the diagonal of the dimer (indicated by the purple line in **a**). Both sections reveal no energy-dependent phase shift and no increased amplitude of the interference pattern around zero bias voltage. Image parameters are (**a**) V = -31 mV, I = 0.2 nA. Spectroscopy parameters are (**b-c**) $V_{SP}$ = -31 mV, $I_{SP}$ = 0.25 nA.

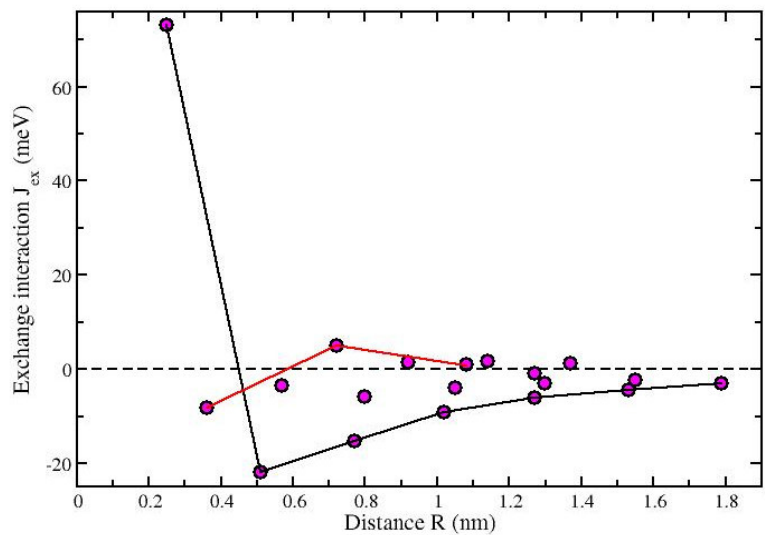

**Supplementary Figure S5. Calculated exchange interaction as function of distance between iron atoms.** While the impurity position below the surface is fixed to the fifth monolayer, the exchange interaction is calculated for various interatomic distances. The black line connects the atoms sitting along the diagonal, the [01$\bar{1}$] direction, which are characterized by strong magnetic exchange interactions. The red line highlights the [001] direction.

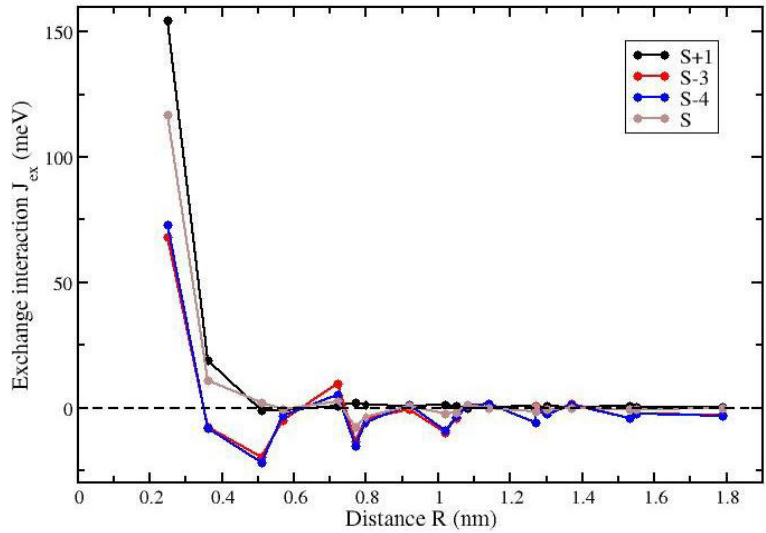

**Supplementary Figure S6. Calculated exchange interaction for impurity positions on top, within and below the surface.** Interestingly, once the impurities are put on top of the surface (S+1), i.e. adatom location, or embedded in the surface (S) the magnetic exchange interactions decay strongly with respect to interatomic distances. As soon as the impurities sit at fourth layer below the surface (S-3) the long-range behaviour with a strong directional anisotropy is observed.

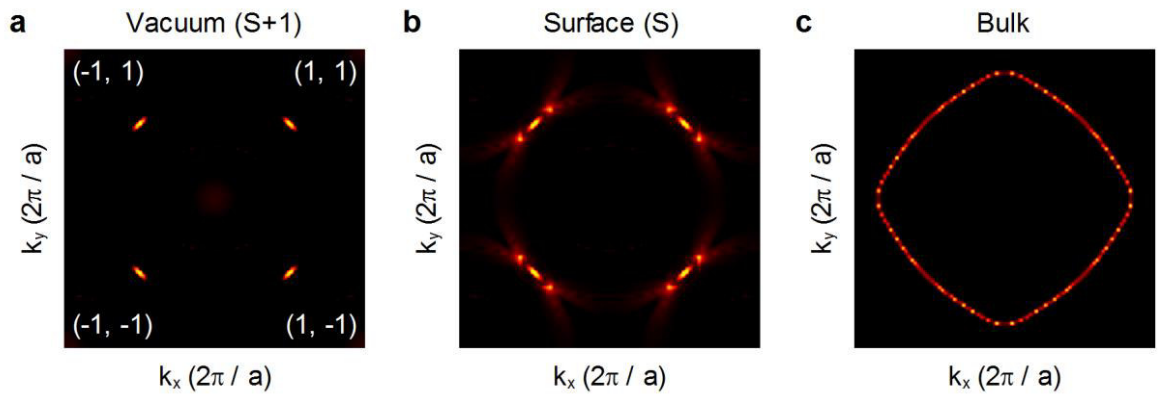

**Supplementary Figure S7. Two-dimensional spectral functions and Fermi energy contours.** The energy contours are highlighted by the electronic occupation via the two-dimensional spectral functions, projected on the vacuum layer above the surface (S+1), on the surface (S) and in bulk (**a-c**). One observes how the bulk energy contour with a square-like shape disappears once on the surface. This explains the large magnetic exchange interactions observed when the impurities are buried below the surface since the flat regions of the energy contours create a strong focusing along the diagonals while on the surface, the flat regions disappear and thereby the magnetic exchange interaction becomes weak.

## Supplementary Tables

| Distance R (nm) | Δ (mV) | φ (π) | ε$_k$ (mV) |
|---|---|---|---|
| 0.51 | no Kondo effect | | |
| 0.57 | 16.5 ± 1.5 | 1.1 ± 0.2 | 0.9 ± 0.9 |
| 0.72 | 15.2 ± 2.0 | 1.2 ± 0.2 | 3.2 ± 1.0 |
| 0.77 | 22.4 ± 3.0 | 1.3 ± 0.1 | -3.9 ± 1.0 |
| 1.02 | 14.0 ± 1.5 | 1.9 ± 0.1 | -0.1 ± 0.7 |
| ∞ | 11.7 ± 3.0 | | |

**Supplementary Table S1. Summary of the Fit parameter determined for different Fe dimers in the fifth monolayer below the Cu(100) surface.** For comparison the resonance width for a single fifth monolayer Fe impurity $\Delta_0$ (R = ∞), averaged over measurements with different tips and spatial positions, is also listed.

## Supplementary Note 1

**Extracting the dimer geometry.** Iron dimer configurations can be identified by their topographic signature [16, 17]. In case where the topographic signature is symmetric both impurities are located within the same monolayer below the surface. Supplementary Fig. S1a shows a topography of two neighbouring Fe atoms. Comparing this with results found for single Fe impurities yields that they are located in the fifth monolayer [17]. The centre of each individual standing-wave pattern, marked by the red dots, reveals a displacement of 0.18 nm along the [010] direction and 0.54 nm along the [001] direction. These lengths correspond to an atomic configuration with interatomic distance of R = 0.57 nm, which is depicted in Supplementary Fig. S1b.

## Supplementary Note 2

**Analysing the spectroscopic signature of a dimer.** In contrast to magnetic atoms and molecules on noble metal surfaces single magnetic impurities embedded below a metal surface induce a complex signature ΔdI/dV(x, y, V) which depends on the lateral tip position (x, y). The main idea for the description of bulk impurities is the separation of the electron propagation, determined by the host properties, and the scattering characteristics of the impurity, which is strongly affected by the Kondo effect and its characteristic energy scale [28]. The same idea can be expanded to a dimer geometry. Here both impurities induce a signature at the surface. As both impurities are Fe atoms and located in the same monolayer below the surface the scattering characteristics of each impurity is equal. However, the characteristic energy scale may change compared to an isolated Fe atom due to magnetic interaction. In general it is complicated to extract microscopic information for a dimer as the spectroscopic signature of a single Kondo impurity, especially the line shape, strongly depends on the distance to each impurity. However, for high symmetry positions of the tip meaning that the distance to both impurities is equal or one impurity is much further away than the other, the same fit formula as for the single impurity case can be used to describe the spectral signature of two interacting impurities (see below). For all other lateral tip positions this model cannot be used as the distance to both impurities is not the same.

**Phenomenological fit formula.** We use a fit formula based on a phenomenological form of the Kondo resonance found by Frota and colleagues [29, 30] which correctly describes Kondo features for a single impurity such as line-shape [31] and more dramatically the phase shift caused by the resonance fairly well in both experiment and many-body calculations [16, 17, 32, 33]. Frota's phenomenological line shape has the following functional form

$$\frac{\Delta dI}{dV}(V) = a \cdot Im\left[-ie^{i\phi}\sqrt{\frac{i0.39\Delta}{eV - \varepsilon_k + i0.39\Delta}}\right] + b \cdot V + c$$

Where $\varepsilon_k$ is the position of the Kondo resonance. The half width at half maximum $\Delta$ of the resonance width is proportional to the Kondo temperature $T_k$, whereas the exact relation is still an open issue. Here $\Delta = 3.7\ T_k$, is used which results from numerical renormalization group (NRG) calculations [32]. A linear voltage slope and an offset are added to account for additional and approximately energy-independent background scattering processes. Error bars resemble the 95 % confidence interval of the parameter and are obtained by the non-linear fit of equation.

As the Kondo resonance cause a phase shift at the Fermi energy the observed signature strongly depends on the distance to the impurity. The line shape of the curve is determined by the line-shape parameter $\phi$, which results for even multiples of $\pi$ ($\phi = 0\pi, 2\pi,...$) in a dip and for odd multiples of $\pi$ ($\phi = \pi, 3\pi,...$) in a peak. Other values cause asymmetric line shapes.

In Supplementary Table S1 the resulting fit parameters, the HWHM of the resonance $\Delta$, the line-shape parameter $\phi$ as well as the resonance position $\varepsilon_k$ are summarized for the different dimer configurations presented in Figure 2. As for single Kondo impurities [17] the line-shape parameter $\phi$ increases for larger lateral tip positions. This behaviour can be seen as a consistency check for the scattering model of two neighbouring atoms based on the separation of the scattering properties and the electron propagation.

**Supplementary Note 3**

**Disappearance of the Kondo effect for a compact dimer configuration.** While for the dimer configurations, with R = 0.57 nm, R = 0.72 nm, R = 0.77 nm and R = 1.02 nm the Kondo effect is still present but with an increased resonance width, for the dimer configuration with R = 0.51 nm no Kondo fingerprints are observed. In Supplementary Fig. S3 single spectra measured at high symmetry points are shown. Spectra taken in the middle of the interference pattern as well as spectra taken at the border reveal a nearly linear voltage dependence and no signatures around zero bias at a sample temperature of 6 K. The small peak for a tip position direct at the centre of the interference pattern (Supplementary Fig. S3c) we do not attribute to Kondo physics as the amplitude is too small and also its position is not at the Fermi energy. More likely the broad feature in the spectroscopy is related to the band structure of the copper crystal. This idea is supported by laterally resolved $\Delta dI/dV$ spectra, depicted in Supplementary Fig. S4. The line cuts do not reveal a phase shift or an enhanced amplitude around zero bias but a small decrease of the lateral size of the interference pattern for positive bias voltages due a change in the wave vector as function of energy. At the centre, where both impurities contribute to the signature, the maxima interfere constructively at around 15 mV leading to broad peak like feature.

**Supplementary Note 4**

**Calculated magnetic exchange interaction.** In Supplementary Fig. S5 the magnetic exchange interactions are shown for the case of impurities embedded in the fifth layer below the surface. Please note, that the surface atomic layer is labelled as S. This means that S-1 is according to the manuscript the second monolayer, S-2 the third monolayer and so on. The resulting exchange interactions are strongly oscillating between ferromagnetic and antiferromagnetic coupling. Note that depending on the probed direction, these interactions can be strongly anisotropic. Along the diagonal the magnetic exchange interactions seem to decay more slowly in comparison to other directions.

In Supplementary Fig. S6 are depicted the magnetic exchange interactions for impurities sitting in different layers, adatom location (S+1), embedded in the surface layer (S) on buried below the surface (S-3) and (S-4). Interestingly as soon as the impurities are rather close to the surface the strength of the magnetic exchange interactions become much smaller compared to the bulk values.

At large distances this is related to the anisotropy of the Fermi surface. In Supplementary Fig. S7 we show the two-dimensional Bloch spectral function at the Fermi energy projected on the vacuum layer above the surface (S+1), on the surface layer (S), and on a bulk layer. This allows us to visualize the Fermi energy contours highlighted by the electronic occupation. Interesting in the bulk, a square-like shape characterizes the energy contour with extremely flat regions. One thus would expect a strong focusing effect of the electronic beams after scattering at a bulk impurity along the directions perpendicular to the flat sides, leading to strong magnetic exchange interactions. This bulk-like contour survive when getting closer to the surface which explains the squared –like contours in the main manuscript of the fifth monolayer (S-4). However on the surface and in the vacuum layer above the surface, these contours have very low occupancy, thus the focusing effect disappears and the magnetic exchange interactions decay strongly with the interatomic distances if the impurities sit in the surface or on top of the surface layer.